# ZMP-SAPT: DFT-SAPT using *ab initio* Densities

A. Daniel Boese[1,a)] and Georg Jansen[2]

[1] *Institute of Chemistry, Physical and Theoretical Chemistry, University of Graz, Heinrichstrasse 28/IV, 8010 Graz, Austria*

[2]*Faculty of Chemistry, University of Duisburg-Essen, Universitätsstraße 5, 45117 Essen, Germany*

Symmetry Adapted Perturbation Theory (SAPT) has become an important tool when predicting and analyzing intermolecular interactions. Unfortunately, DFT-SAPT, which uses Density Functional Theory (DFT) for the underlying monomers, has some arbitrariness concerning the exchange-correlation potential and the exchange-correlation kernel involved. By using *ab initio* Brueckner Doubles densities and constructing Kohn-Sham orbitals via the Zhao-Morrison-Parr (ZMP) method, we are able to lift the dependence of DFT-SAPT on DFT exchange-correlation potential models in first order. This way, we can compute the monomers at the Coupled-Cluster level of theory and utilize SAPT for the intermolecular interaction energy. The resulting ZMP-SAPT approach is tested for small dimer systems involving rare gas atoms, cations, and anions and shown to compare well with the Tang-Toennies model and coupled cluster results.

a) Author to whom correspondence should be addressed. Electronic mail: Adrian_Daniel.Boese@uni-graz.at

## I. INTRODUCTION

Intermolecular interactions are still one of the big challenges to modern quantum chemistry. To accurately describe these interactions, modern quantum chemical methods need to be applied. Here, dispersion-corrected density functional theory (DFT+D) may not be enough, as a closer look at the performance of density functionals for hydrogen bonds reveals.[1,2]

From the currently available successful post-Hartree-Fock methods, second-order Møller-Plesset perturbation theory (MP2) is regarded as one of the simplest. The accuracy of MP2 is higher than that of DFT+D for hydrogen bonds.[1] However, for molecules, where van der Waals interactions become more important, it may not be the method of choice as demonstrated by a systematic overestimation of the magnitude of interaction energies between molecules with delocalized $\pi$–systems.[3-5] The accuracy of coupled-cluster theory using single, double and perturbative triple excitations (CCSD(T)) is much higher, but this method is computationally rather expensive, despite recent progress.[6-12] Another alternative, symmetry perturbation theory (SAPT)[13] in its DFT-SAPT variant is using density functional theory only for the monomers.[14-19] SAPT is often used to describe intermolecular interactions. It will start to fail as soon as covalent bonding and significant charge-transfer is involved, however it yields good interaction energies for hydrogen bonds[20] and excellent results when van der Waals interactions are calculated.[21] SAPT using MP2 and coupled-cluster theory at the CCSD, and, for the dispersion contribution, CCD+ST(CCD) levels is possible,[13,22-25] but has a prohibitive scaling which is about as steep than that of CCSD(T). DFT-SAPT in its density fitting variant,[18,26] however, is rather fast and can be used for medium to large sized systems.

The first order single determinant based Hartree-Fock (HF)-SAPT and DFT-SAPT contributions can be computed from just the orbitals that can be obtained from coupled single-electron wave function equations where only the so-called exchange-correlation (xc) potential constitutes an unknown quantity and which therefore needs to be approximated. A knowledge of the exact xc potential of Kohn-Sham theory would yield the exact first-order electrostatic interaction energy in this framework (in the complete basis set limit), while the first-order exchange (aka exchange-overlap) contribution can be hoped to be well-approximated in this case.[15,16] Higher order interaction energy contributions require solution of (time-dependent) coupled-perturbed Kohn-Sham (or Hartree-Fock) equations and thus knowledge of the second and higher order functional derivatives of the xc functional.. The total interaction energy $\Delta E_{int}$ in DFT-SAPT is conventionally computed by the sum

$$\Delta E_{int} = E_{el}^{(1)} + E_{exch}^{(1)} + E_{ind}^{(2)} + E_{exch-ind}^{(2)} + E_{disp}^{(2)} + E_{exch-disp}^{(2)} + \delta(HF) \qquad (1)$$

, which is composed of the first-order electrostatic and exchange (*aka* exchange-overlap) terms, the second-order induction, exchange-induction, dispersion, exchange-dispersion and the $\delta(HF)$ term. The use of the latter term is somewhat debated, as it emerges as the difference of a simple Hartree-Fock interaction energy calculation and a corresponding HF-SAPT calculation in order to estimate the higher-order induction and exchange-induction terms (note that all terms contained in parentheses are determined from HF orbitals and coupled-perturbed HF amplitudes):[27-29]

$$\delta(HF) = E_{int}^{HF} - \left(E_{el}^{(1)} + E_{exch}^{(1)} + E_{ind}^{(2)} + E_{exch-ind}^{(2)}\right) \tag{2}$$

While the basis set superposition error (BSSE) plaguing supermolecular calculations of interaction energies has to be corrected for the determination of $E_{int}^{HF}$ by the counterpoise-correction scheme,[30] the SAPT interaction energy is zeroth-order counterpoise-corrected by construction.

For computing intermolecular interactions, SAPT is often the method of choice, with the problem that the underlying density functional is still introducing some arbitrariness from the xc potential as well as the xc kernel.

Through comparison to benchmark-quality Coupled-Cluster based SAPT, Korona identified asymptotically-corrected[31-33] variants of the PBE0[34,35] and B3LYP[36-38] hybrid xc potentials as good choices for DFT-SAPT calculations.[22] While interaction energies obtained with them are very useful for vibrational-rotational tunneling spectroscopy of van der Waals complexes,[39,40] they can certainly not be considered as spectroscopically accurate.[41,42] On the other hand, since Kohn-Sham DFT establishes an invertible connection between the xc potential and the electron density, the density can be used to construct the xc potential.[43-46] In the two-electron case, an exact Hylleraas density can be inverted to yield an essentially exact xc potential, a possibility which has been exploited for SAPT calculations of the Helium dimer.[47] Limited tests for SAPT calculations with *ab initio* density derived xc potentials have been performed on $Ne_2$, $Ar_2$, and NeAr, using the Zhao-Morrison-Parr method.[48] Here, we employ the Zhao-Morrison-Parr (ZMP) method[46] to construct xc potentials from *ab initio* densities for a range of test systems including ions (please note that the optimized effective potential (OEP) method provides another route to almost exact xc potentials which are derived from ab initio approaches such as Moller-Plesset or coupled cluster theory'[49,50]). These are in turn used in the SAPT formalism to calculate the interaction energies between the monomers, arriving at what will be termed ZMP-SAPT.

## II. Computational Details and Methodology

For the DFT-SAPT and CCSD(T) calculations, we used the MOLPRO suite of programs.[51] For constructing the ZMP potentials based on Brueckner Doubles (BD) calculations which are based on the CCSD formalism, we utilized CADPAC[52] and a TZ2P basis set.[53] In all DFT-SAPT and CCSD(T) calculations, Dunning's basis sets aug-cc-pVXZ with diffuse functions were employed.[54-56] When extrapolating the aug-cc-pV(X,Y)Z basis sets, for the correlation energy (or in the case of DFT-SAPT dispersion and exchange-dispersion parts), the standard $X^{-3}$ formula[57] and for the Hartree-Fock contribution, the $e^{(-9\sqrt{X})}$ formula[58,59] have been used. Together with the ZMP method, several functionals have been tested, namely PBE,[34] PBE0[35] and the asymptotically corrected PBE0AC[31,60] which is based on the LB94 long-range potential[32] and the GRAC connection scheme.[33] The shift parameters (0.229 Hartree for He, 0.200 Hartree for Ne, 0.137 Hartree for Ar, 0.116 Hartree for Li$^+$, 0.097 Hartree for Na$^+$, 0.402 Hartree for F$^-$, 0.269 Hartree for Cl$^-$) required for the asymptotic correction were determined as the sum of the (positive) ionization potential and the (negative) energy of the highest occupied molecular orbital. Both were determined at the PBE0/aug-cc-pV5Z level of theory. For the rare gas atoms, they are by 0.002-0.005 Hartree (i.e. about 2%) smaller than the shift parameters used in references 41 and 42.

The underlying kernel for the $2^{nd}$-order DFT-SAPT contribution was the pure ALDA kernel[61] when semi-local xc potentials (PBE,ZMP) were used and a hybrid ALDA kernel[62] for the hybrid xc potentials (PBE0,PBE0AC). As it is customary in SAPT, the S² approximation has been invoked for the second-order exchange-induction and exchange-dispersion contributions, while it has been avoided for the first-order exchange contribution.[13,63,64]

Although the ZMP equations have been introduced elsewhere,[46] we will give a short overview for the closed-shell case. The Kohn-Sham orbital equations read:

$$\hat{h}_{eff}\phi_a(\vec{r}) = \varepsilon_a\phi_a(\vec{r}) = \left(\hat{h}(\vec{r}) + v_J(\vec{r}) + v_{xc}(\vec{r})\right)\phi_a(\vec{r})$$
$$= \left(\hat{h}(\vec{r}) + v_J(\vec{r}) + \frac{\delta F_{xc}(\rho(\vec{r}))}{\delta\rho(\vec{r})}\right)\phi_a(\vec{r}) \tag{3}$$

with the orbitals $\phi_a$ and the effective Hamiltonian $\hat{h}_{eff}$. This Hamiltonian consists of the one-electron operator $\hat{h} = -\frac{1}{2}\nabla^2 + v_{eN}(\vec{r})$, the electronic coulomb potential $v_J(\vec{r}) = \int\frac{\rho(\vec{r}')}{|\vec{r}-\vec{r}'|}$, and the xc potential $v_{xc}(r)$ which is the derivative of the exchange-correlation functional $F_{xc}(\rho(r))$. The density $\rho(\vec{r})$, which integrates to the electron number $N$, is given by $\rho(\vec{r}) = \sum_a|\phi_a(\vec{r})|^2$, with $a$ being the index of the occupied orbitals, assuming orthonormality of the orbitals:

$$\int\phi_a^*(\vec{r})\phi_b(\vec{r})dr = \delta_{ab} \tag{4}$$

To derive the ZMP equations through the constrained-search minimization of the kinetic energy functional for the 'non-interacting electrons' of Kohn-Sham theory one introduces the constraint:

$$\frac{1}{2} \int \int \frac{[\rho(\vec{r}) - \rho_{exact}(\vec{r})][\rho(\vec{r}') - \rho_{exact}(\vec{r}')]}{|\vec{r} - \vec{r}'|} d\vec{r} d\vec{r}' = 0 \tag{5}$$

where $\rho_{exact}(\vec{r})$ is the exact ground state density, assumed to be known. Taking the functional derivative of this and plugging it with a Lagrangian multiplier $\lambda$ as a substitute for $v_{xc}(\vec{r})$ into equation (3) leads to:

$$\left( \hat{h}(\vec{r}) + \left( 1 - \frac{1}{N} \right) v_J^\lambda(\vec{r}) + v_{xc}^\lambda(\vec{r}) \right) \phi_a^\lambda(\vec{r}) = \varepsilon_i \phi_a^\lambda(\vec{r}) \tag{6}$$

where the Coulomb operator has been corrected by the Fermi-Amaldi factor $\left( 1 - \frac{1}{N} \right)$ [65] in order to ensure the proper $-\frac{1}{r}$ asymptotic behavior of the resulting xc potential

$$v_{xc}(\vec{r}) = \lim_{\lambda \to \infty} \left[ \lambda \int \frac{\rho^\lambda(\vec{r}') - \rho_{exact}(\vec{r}')}{|\vec{r} - \vec{r}'|} d\vec{r}' - \frac{1}{N} \int \frac{\rho^\lambda(\vec{r}')}{|\vec{r} - \vec{r}'|} d\vec{r}' \right] \tag{7}$$

and thus guaranteeing that the orbital energy of the highest occupied orbital coincides with the first ionization potential. The above equations can be solved using post Hartree-Fock densities (in our case BD/TZ2P) as "exact" densities on a numerical grid. While $\lambda$ formally has to be extrapolated to infinity to obtain $v_{xc}(\vec{r})$, a value of 900 has been found reasonable for the purpose at hand.[66,67]

### III. Results and Discussion

#### A. From ZMP to ZMP-SAPT

As we have mentioned in the last section, all ZMP xc potentials are calculated on a grid. Grids usually considered for molecules are direct products of Euler-Maclaurin[68] or other radial grids with Lebedev angular grids.[69-72] For atoms because of their symmetrical shape, Gauss-Legendre spherical products are more convenient. The ghost-atoms in the DFT-SAPT calculation, however, include also grid points on the position of the second atom. Furthermore, for obtaining numerically correct integrals involving the xc potentials, a large amount of xc potential grid points has to be used, whereas the ZMP calculations only converge if the calculations are done on a rather small number of xc potential grid points[73] (see Fig. 1).

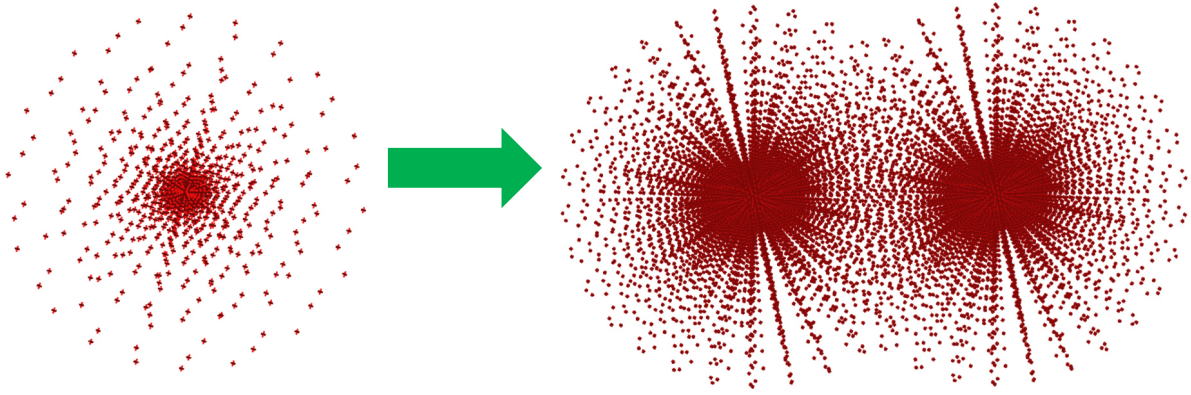

FIG. 1. The technical difficulty concerning the grid points. To the left are the grid points of the ZMP grid of one argon atom on which the xc potential has been calculated. To the right are the grid points needed by the DFT-SAPT calculation when the argon-argon interaction energy is calculated.

For atoms, this is an easy task to solve, as we can do a simple 1D-least squares fit to obtain the additional grid points, like done in Fig. 2. By using this method, 18946 points have been fitted from 1664 points for the Helium dimer, 31728 from 2000 points for the Neon dimer, and 39298 from 1570 for the Argon dimer. Similar grid sizes of a few ten thousand points were employed for the other dimers involving ions.

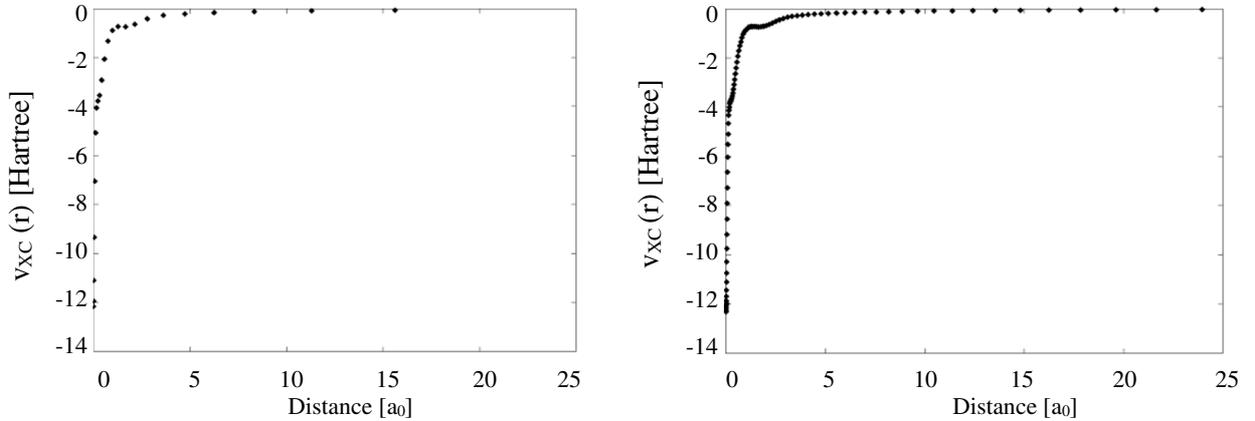

FIG. 2. The xc potential in dependence on the distance of the argon atom, as shown in Fig. 1. To the left is the ZMP grid, to the right the grid (where the additional points have been obtained by a 1-D-least-squares fit) from which the DFT-SAPT energy is computed.

With this in mind, we can now calculate ZMP-SAPT interaction energies for closed-shell-atoms and ions in order to test the performance of the new method.

## B. Binding Characteristics of the Systems investigated

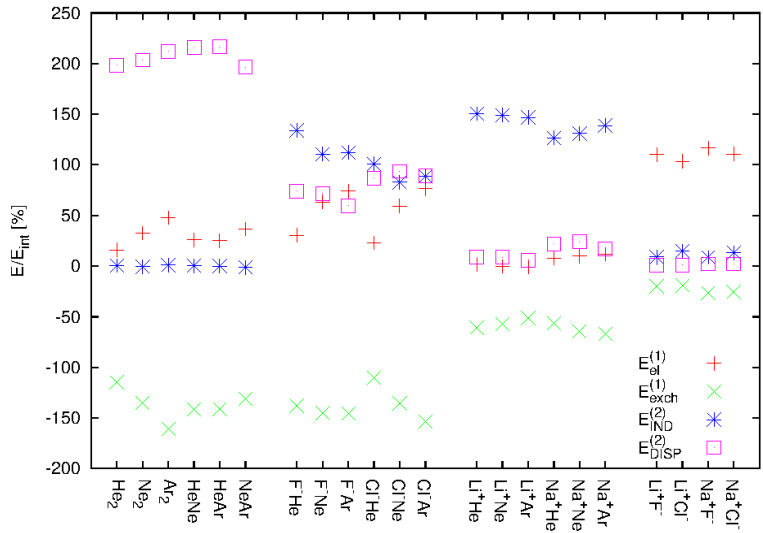

FIG. 3: Individual SAPT interaction energies of all investigated systems.

Before studying the performance of ZMP-SAPT in comparison to other versions of SAPT and to CCSD(T) in detail let us investigate the characteristics of the binding between the various atoms and ions considered here. Fig. 3 displays the SAPT contributions $E_{el}^{(1)}$, $E_{exch}^{(1)}$, $E_{IND}^{(2)} = E_{ind}^{(2)} + E_{exch-ind}^{(2)} + \delta(HF)$, and $E_{DISP}^{(2)} = E_{disp}^{(2)} + E_{exch-disp}^{(2)}$ as percental contributions to the sum $E_{int}$ of all first- and second-order terms for the equilibrium distances of each respective dimer. Unsurprisingly, in case of the rare gas dimers the dispersion term causes the weak, non-covalent binding, with the repulsive first-order exchange contribution canceling somewhat more than half of the dispersion contribution at the van der Waals minimum. The strong ionic binding between anions and cations, on the other hand, is nearly exclusively due to electrostatic interactions, as to be expected, with minor contributions of the induction and first-order exchange terms, while dispersion interactions are negligible here. Interestingly, the binding between cations and rare gas atoms is dominated by the induction contribution: the small and „hard" cations can strongly polarize neutral atoms and also induce a certain amount of charge transfer from their interaction partner (note that without further measures[74] SAPT includes charge transfer in the induction term). Roughly one third of the resulting attraction is cancelled by the first-order exchange term. The most interesting cases, however, are found for interactions between anions and rare gas atoms: here all three, electrostatic, induction, and dispersion contributions are significant for the binding, in the case of chloride-argon each contributing about the same amount. Here, induction and dispersion are significant since both, the anion and the rare gas atom can be polarized with ease. The significance of the electrostatic contribution may be surprising at first sight, yet one should take into account that the nuclear charge of each interaction partner is not completely screened by its electron cloud when the atoms and anions start to interpenetrate. This is resulting in a dominance of the attraction between the nucleus of one atom / anion and the electron cloud of the other anion / atom – at least as long as the nuclei do not come too close. This effect is also visible for the rare gas dimers, where in the case of the argon dimer $E_{el}^{(1)}$ contributes about 50% to the binding.

Summarizing, in the following we cover a wide range of possible interaction characteristics, from dispersion- (rare gas dimers) over induction- (cation - rare gas) to electrostatics-dominated (anion – cation) dimers, including "mixed" interactions (anion – rare gas), where all contributions are significant.

### C. Interaction Energies for Rare Gas Atoms

In Fig. 4 (a-f), we show the interaction potentials of the He$_2$, Ne$_2$, Ar$_2$, HeNe, HeAr and NeAr dimers as obtained with DFT-SAPT/aug-cc-pV6Z for various xc potentials in comparison to the Tang-Toennies model,[75] theoretical as well as to CCSD(T)/aug-cc-pV6Z values.

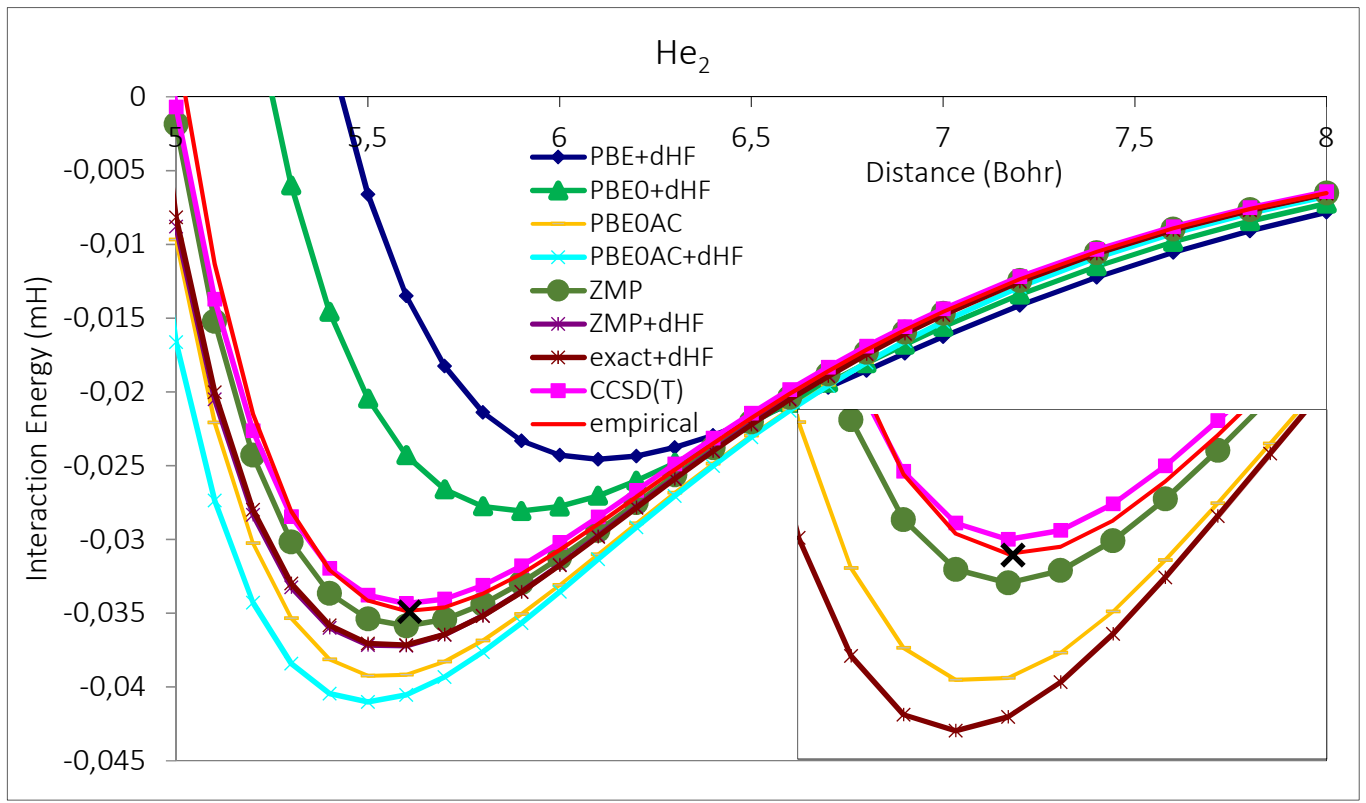

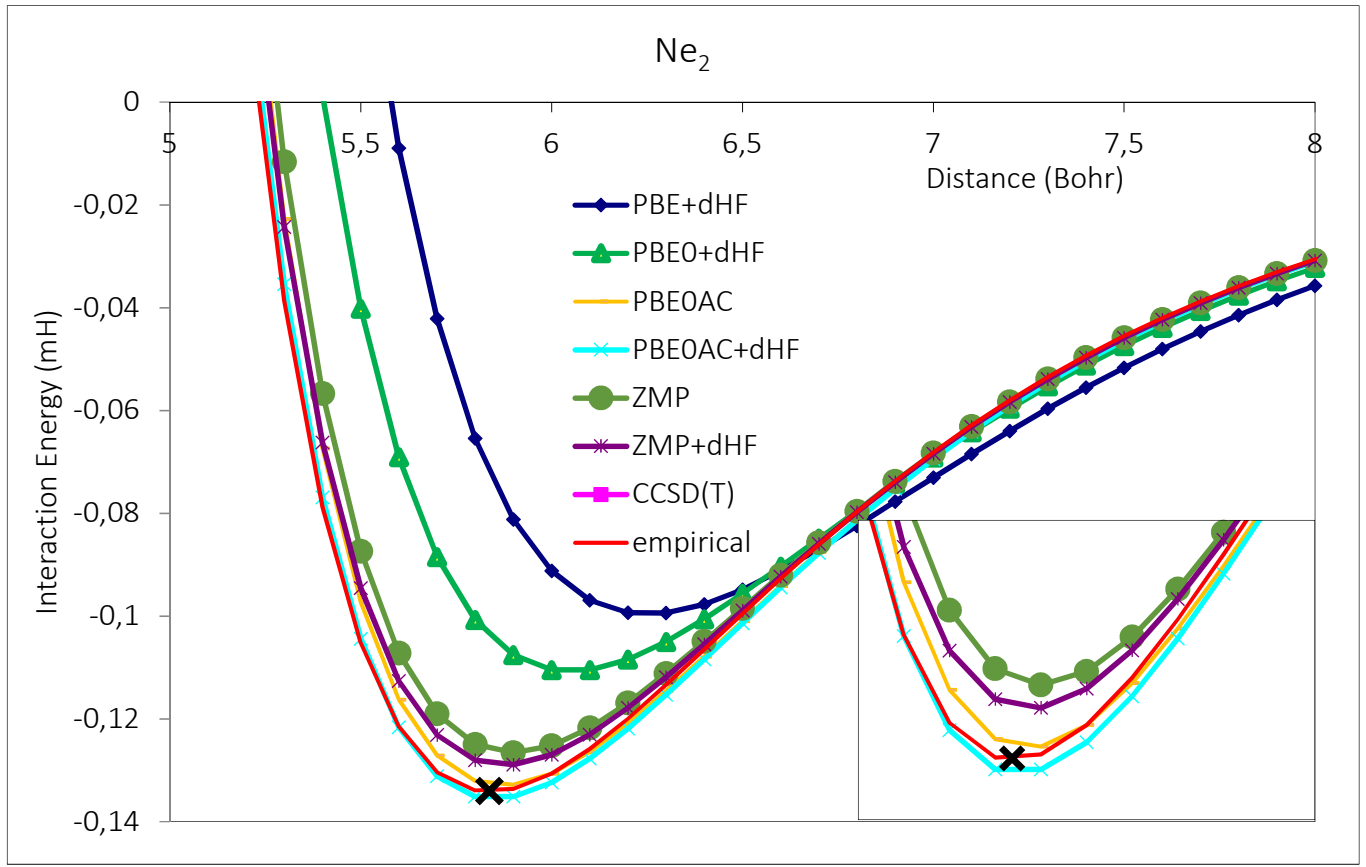

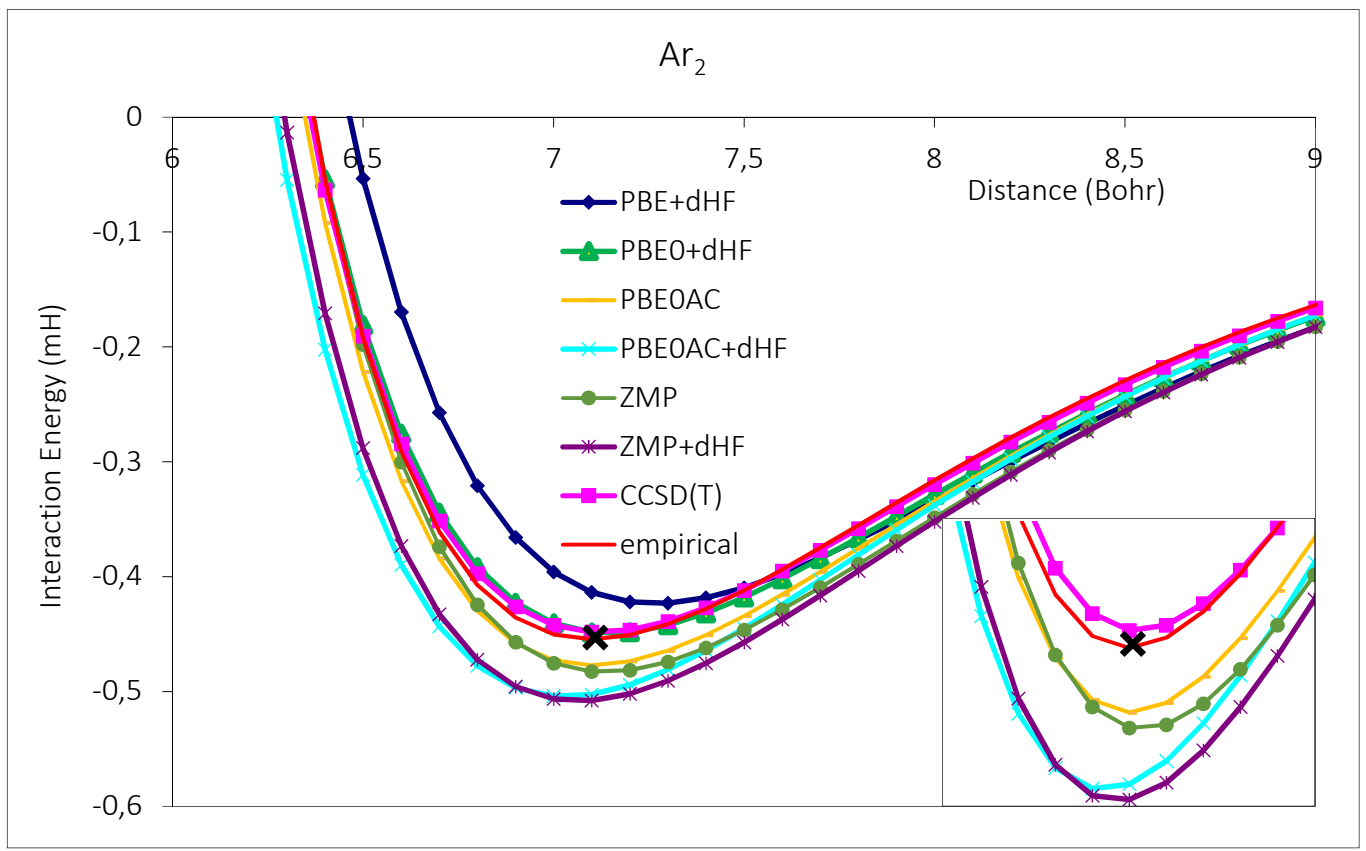

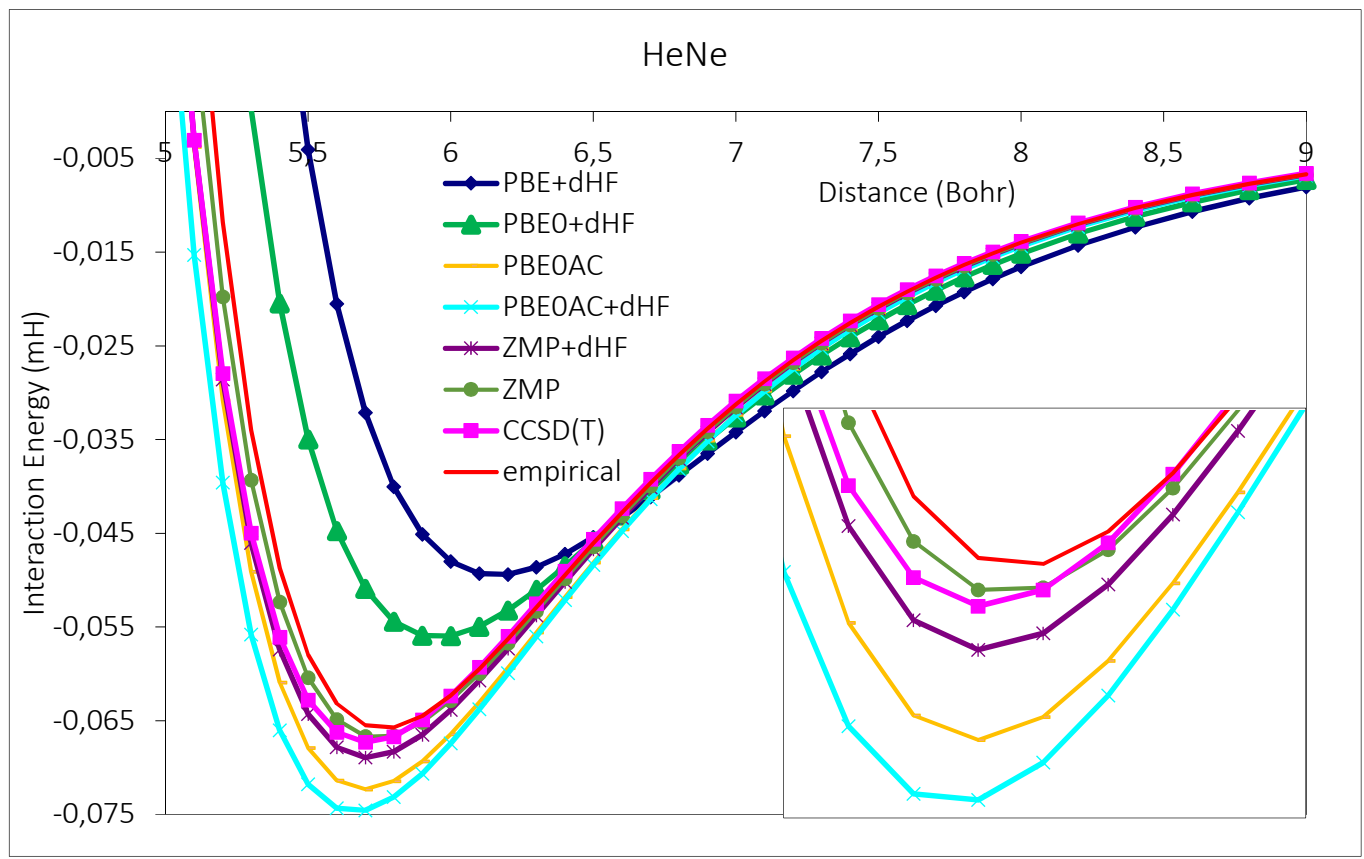

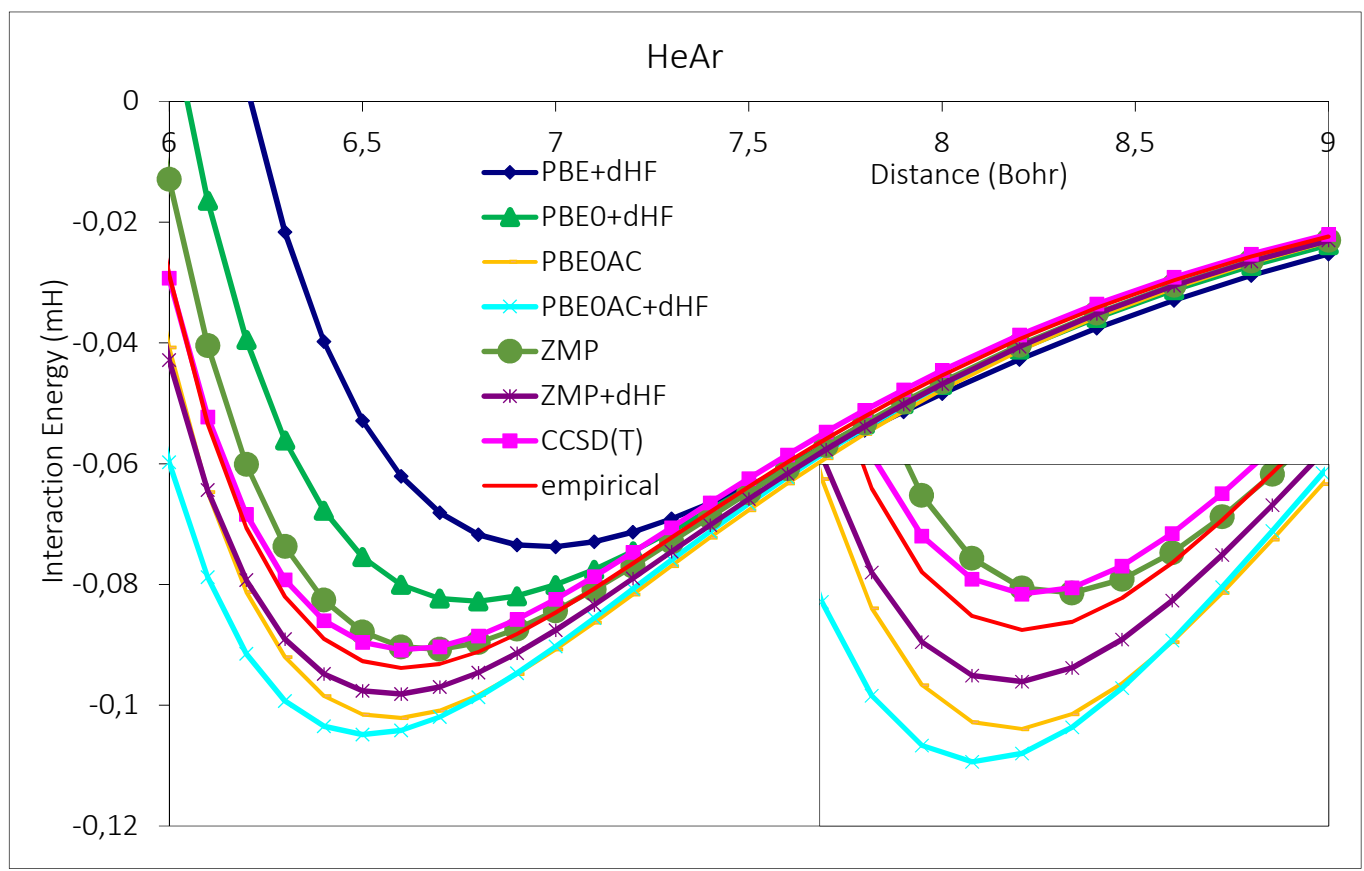

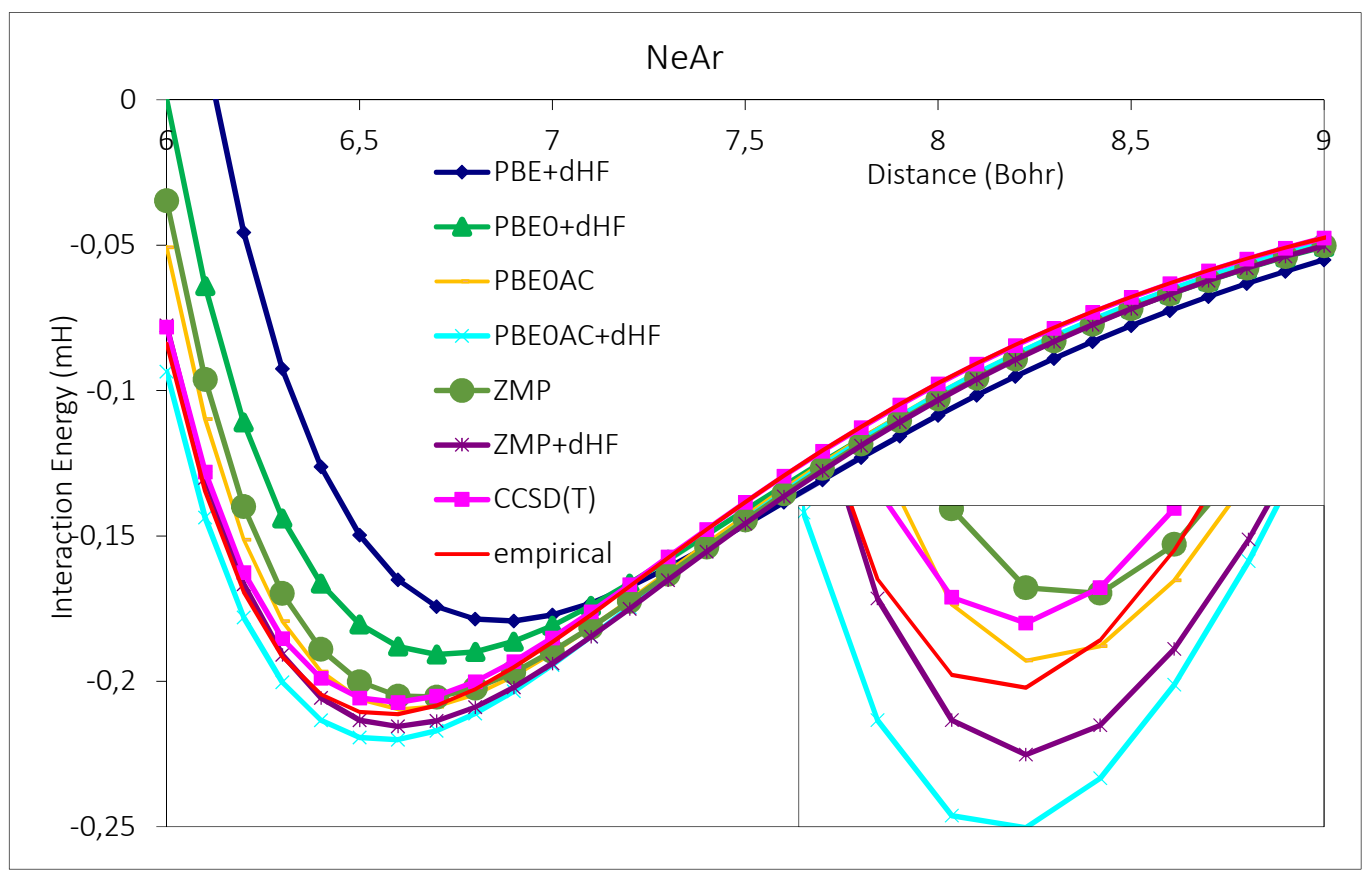

FIG. 4. The potential energy curves of DFT-SAPT with various xc potentials with and without $\delta(HF)$ correction: PBE-SAPT+ $\delta(HF)$ (blue), PBE0-SAPT+ $\delta(HF)$ (green), PBE0AC-SAPT (orange) and PBE0AC-SAPT+ $\delta(HF)$ (turquoise), ZMP-SAPT+ $\delta(HF)$ (purple) and ZMP-SAPT (dark green) in comparison to cp-corrected CCSD(T) (pink), Tang-Toennies values (red)[75], best available theoretical results (black cross)[76-78], together with the curve from the exact xc potential for helium (brown) for the helium dimer (a), neon dimer (b), argon dimer (c), HeNe (d), HeAr (e) and NeAr (f). All BD densities for ZMP xc potential generation have been computed with the TZ2P basis set, whereas the DFT-SAPT and the counterpoise-corrected CCSD(T) curves have been computed with aug-cc-pV6Z.

Except for the case of HeNe, with CCSD(T) at this basis set level the well depths are slightly underestimated compared to the empirical Tang-Toennies data. As summarized in reference 41, achieving the complete basis set (cbs) limit through extrapolation or explicit correlation techniques (such as F12 variants), possibly in combination with inclusion of iterative triple and perturbative quadruple excitations, brings Coupled Cluster results into the spectroscopic accuracy regime, with deviations of fractions of a wavenumber from the most reliable experimental results for the well depths of the rare gas dimers considered here. However, in the following we consider CCSD(T) with a specific basis set as homogeneous fundament for comparison to DFT-SAPT in the same basis set, since for some of the ion-containing dimers (vide infra) accurate empirical potential energy curves are not available. For simplicity, the empirical Tang-Toennies model curves[75] were taken as empirical potential curves for all rare gas dimers.

From Figure 4 it is obvious that the xc potential as obtained from the PBE functional yields strongly underbinding DFT-SAPT interaction potentials for all of the considered rare gas dimers, despite inclusion of the $\delta(HF)$ correction which in general enhances the well depth (vide infra). Replacing PBE with the PBE0 hybrid xc potential with its 25% "exact exchange" contribution lowers the potential wells significantly, such that in the case of $Ar_2$ very good agreement with both, the CCSD(T) and the empirical Tang-Toennies potential curves is obtained. For the remaining five rare gase dimers, however, DFT-SAPT(PBE0)+$\delta(HF)$ is far from satisfactory: here all potential energy curves remain significantly underbinding.

This changes when the asymptotic correction is applied to PBE0: in nearly all cases the potential curve obtained with PBE0AC and inclusion of the $\delta(HF)$ correction is the deepest curve, and in the case of $Ar_2$ (as the only exception) it nearly is. As a consequence, DFT-SAPT(PBE0AC)+$\delta(HF)$ is overbinding the rare gas dimers, as has been noticed before in reference 41 where it was also observed that avoiding the $\delta(HF)$ correction leads to better agreement with benchmark results.

However, the most relevant observation from Figure 4 in the context of the present work is the overall good agreement of the ZMP-SAPT potential curves with the benchmark data: if the $\delta(HF)$ term is omitted they are somewhat too deep with respect to the empirical Tang-Toennies curves in three cases ($He_2$, $Ar_2$, and HeNe) and somewhat too shallow in the remaining cases ($Ne_2$, HeAr, and NeAr). It should be noted that the present ZMP-SAPT results for $He_2$ are extremely close to those obtained previously with an essentially exact xc potential[48] since BD is exact within the given basis set, and that adding a full configuration interaction correction $\delta(FCI)$ to them brings them to excellent agreement with the empirical Tang-Toennies data.[47] Comparing ZMP-SAPT without any correction to CCSD(T) in the same basis set one observes very good agreement for all of the "mixed" dimers (HeNe, HeAr, and NeAr) , a slight overbinding for $He_2$ and a slight underbinding for $Ne_2$. The most significant discrepancies are found for $Ar_2$, where ZMP-SAPT significantly overbinds with respect to CCSD(T).

Table 1: Individual ZMP-SAPT and DFT-SAPT(PBE0AC) contributions (in milliHartree) at the approximate minimum distances (in Bohr) of the neutral dimers.

| ZMP | Min. Dist. | $E^{(1)}_{pol}$ | $E^{(1)}_x$ | $E^{(2)}_{ind}$ | $E^{(2)}_{ind-x}$ | $E^{(2)}_{disp}$ | $E^{(2)}_{disp-x}$ | $E^{(1)}_{tot}$ | $E^{(2)}_{tot}$ | $E_{int}$ |
|---|---|---|---|---|---|---|---|---|---|---|
| He$_2$ | 5.6 | -0.0054 | 0.0387 | -0.0009 | 0.0007 | -0.0691 | 0.0022 | 0.0334 | -0.0671 | -0.0337 |
| Ne$_2$ | 5.9 | -0.038 | 0.160 | -0.033 | 0.034 | -0.253 | 0.013 | 0.121 | -0.240 | -0.118 |
| Ar$_2$ | 7.1 | -0.219 | 0.738 | -0.300 | 0.295 | -1.063 | 0.090 | 0.519 | -0.978 | -0.459 |
| HeNe | 5.7 | -0.0161 | 0.0882 | -0.0074 | 0.0073 | -0.1401 | 0.0058 | 0.0072 | -0.1344 | -0.0623 |
| HeAr | 6.6 | -0.0226 | 0.1278 | -0.0076 | 0.0077 | -0.204 | 0.009 | 0.1052 | -0.1955 | -0.0903 |
| NeAr | 6.7 | -0.071 | 0.255 | -0.081 | 0.084 | -0.406 | 0.025 | 0.184 | -0.378 | -0.194 |
| | | | | | | | | | | |
| PBE0AC | Min. Dist. | $E^{(1)}_{pol}$ | $E^{(1)}_x$ | $E^{(2)}_{ind}$ | $E^{(2)}_{ind-x}$ | $E^{(2)}_{disp}$ | $E^{(2)}_{disp-x}$ | $E^{(1)}_{tot}$ | $E^{(2)}_{tot}$ | $E_{int}$ |
| He$_2$ | 5.6 | -0.0053 | 0.0373 | -0.0008 | 0.0007 | -0.0692 | 0.0020 | 0.0320 | -0.0673 | -0.0353 |
| Ne$_2$ | 5.9 | -0.037 | 0.152 | -0.029 | 0.030 | -0.254 | 0.012 | 0.116 | -0.241 | -0.125 |
| Ar$_2$ | 7.1 | -0.198 | 0.665 | -0.251 | 0.247 | -0.997 | 0.082 | 0.467 | -0.920 | -0.453 |
| HeNe | 5.7 | -0.0157 | 0.0843 | -0.0068 | 0.0067 | -0.1487 | 0.0054 | 0.0686 | -0.1332 | -0.0646 |
| HeAr | 6.6 | -0.0212 | 0.1173 | -0.0067 | 0.0066 | -0.1959 | 0.0075 | 0.0960 | -0.1881 | -0.0921 |
| NeAr | 6.7 | -0.065 | 0.236 | -0.070 | 0.072 | -0.385 | 0.022 | 0.170 | -0.360 | -0.190 |

It is noteworthy that the overbinding of DFT-SAPT for Ar$_2$ with the ZMP xc potential is even slightly more pronounced than with PBE0AC. In a recent study of Shirkov and Sladek (note that two different asymptotic corrections have been investigated in reference 42, in which the label GR-LB refers to the same asymptotic correction as used here, except for slight changes in the shift parameter (vide supra)), the overbinding of DFT-SAPT(PBE0AC) total interaction energies for Ar$_2$ has been attributed to an underestimation of the repulsive first order exchange-overlap contribution $E^{(1)}_{exch}$. And in fact, with the ZMP-SAPT approach $E^{(1)}_{exch}$ is systematically more repulsive (by $5-11\%$), as shown in Table 1 for distances close to the equilibrium distances. While this is somewhat counteracted by a more attractive first-order electrostatic interaction contribution $E^{(1)}_{el}$ (also by up to 11%), the resulting total first-order interaction energies $E^{(1)}_{tot}$ become also systematically more repulsive with ZMP-SAPT. The total second-order interaction energies $E^{(2)}_{tot}$, on the other hand, remain nearly constant when replacing the PBE0AC with the ZMP xc potential – with exception of NeAr and Ar$_2$, where they become more attractive by about 0.02 and 0.06 mH, respectively. This reflects corresponding trends in the dispersion energy $E^{(2)}_{disp}$ as the main contributor to $E^{(2)}_{tot}$. As a consequence, while the magnitudes of most $E_{int}$ decrease upon replacing PBE0AC with ZMP, they actually slightly increase for NeAr and Ar$_2$.

Table 2: Well depths D$_e$ (in milliHartree) for several closed-shell diatomic dimers, comparing DFT-SAPT with cp-corrected CCSD(T). All calculations were performed with an aug-cc-pV6Z basis set, except the lithium

containing molecules (aug-cc-pVQZ) and sodium containing ones (aug-cc-pV5Z). The last row contains the percentage root-mean-squared deviation (RMS) compared to the CCSD(T) values. Please note that the dissociation of $Li^+F^-$, $Li^+Cl^-$, $Na^+F^-$, and $Na^+Cl^-$ into ions is considered and not into radicals. All minima were determined with an Akima 1D Spline.

| Dimer | DFT-SAPT | | | | | | CCSD(T) |
|---|---|---|---|---|---|---|---|
| | PBE+$\delta(HF)$ | PBE0+$\delta(HF)$ | PBE0AC | PBE0AC+$\delta(HF)$ | ZMP | ZMP+$\delta(HF)$ | |
| $He_2$ | 0.0228 | 0.0263 | 0.0370 | 0.0386 | 0.0338 | 0.0351 | 0.0325 |
| $Ne_2$ | 0.093 | 0.103 | 0.125 | 0.128 | 0.119 | 0.121 | 0.123 |
| $Ar_2$ | 0.400 | 0.426 | 0.453 | 0.479 | 0.460 | 0.484 | 0.427 |
| HeNe | 0.0459 | 0.0524 | 0.0679 | 0.0702 | 0.0627 | 0.0645 | 0.0634 |
| HeAr | 0.0738 | 0.0828 | 0.0963 | 0.1049 | 0.0907 | 0.0982 | 0.0908 |
| NeAr | 0.168 | 0.179 | 0.198 | 0.208 | 0.194 | 0.203 | 0.196 |
| $F^-He$ | 0.080 | 0.212 | 0.275 | 0.311 | 0.315 | 0.374 | 0.337 |
| $F^-Ne$ | 0.428 | 0.643 | 0.682 | 0.737 | 0.796 | 0.884 | 0.840 |
| $F^-Ar$ | 3.10 | 3.75 | 3.70 | 3.91 | 4.33 | 4.66 | 4.00 |
| $Cl^-He$ | 0.108 | 0.160 | 0.169 | 0.213 | 0.187 | 0.249 | 0.194 |
| $Cl^-Ne$ | 0.400 | 0.470 | 0.439 | 0.522 | 0.496 | 0.610 | 0.526 |
| $Cl^-Ar$ | 2.23 | 2.41 | 2.30 | 2.52 | 2.61 | 2.88 | 2.30 |
| $Li^+He$ | 2.73 | 2.76 | 2.77 | 2.97 | 2.67 | 2.86 | 2.95 |
| $Li^+Ne$ | 4.19 | 4.19 | 4.12 | 4.43 | 4.02 | 4.32 | 4.48 |
| $Li^+Ar$ | 10.07 | 10.17 | 9.66 | 10.51 | 9.92 | 10.82 | 10.71 |
| $Na^+He$ | 1.430 | 1.430 | 1.501 | 1.486 | 1.365 | 1.361 | 1.511 |
| $Na^+Ne$ | 2.34 | 2.23 | 2.37 | 2.38 | 2.23 | 2.23 | 2.43 |
| $Na^+Ar$ | 5.84 | 5.82 | 6.07 | 5.91 | 6.07 | 5.95 | 6.15 |
| $Li^+F^-$ | 295.3 | 291.6 | 283.6 | 292.5 | 285.5 | 294.4 | 293.4 |
| $Li^+Cl^-$ | 240.7 | 242.7 | 239.9 | 244.2 | 239.7 | 243.6 | 244.5 |
| $Na^+F^-$ | 247.0 | 245.9 | 248.5 | 245.7 | 248.9 | 246.1 | 245.5 |
| $Na^+Cl^-$ | 209.0 | 209.8 | 215.2 | 210.0 | 214.0 | 209.0 | 210.8 |
| RMS % | 25.5 | 13.0 | 8.9 | 7.7 | 6.4 | 11.0 | |

Table 3: Equilibrium distances $R_e$ (in Bohr) for several closed-shell diatomic dimers, comparing DFT-SAPT with cp-corrected CCSD(T). All calculations were performed with an aug-cc-pV6Z basis set, except the lithium containing molecules (aug-cc-pVQZ) and soldium containing ones (aug-cc-pV5Z). The second last row denotes the root-mean-squared deviation (RMS) compared to the CCSD(T) values and the last row the corresponding percentage RMS deviation. All minima were determined with an Akima 1D Spline.

| Dimer | DFT-SAPT | | | | | | CCSD(T) |
|---|---|---|---|---|---|---|---|
| | PBE+$\delta(HF)$ | PBE0+$\delta(HF)$ | PBE0AC | PBE0AC+$\delta(HF)$ | ZMP | ZMP+$\delta(HF)$ | |
| He$_2$ | 6.15 | 5.93 | 5.57 | 5.53 | 5.63 | 5.58 | 5.64 |
| Ne$_2$ | 6.28 | 6.09 | 5.90 | 5.88 | 5.94 | 5.91 | 5.88 |
| Ar$_2$ | 7.30 | 7.18 | 7.14 | 7.06 | 7.16 | 7.09 | 7.16 |
| HeNe | 6.24 | 6.02 | 5.76 | 5.72 | 5.80 | 5.77 | 5.77 |
| HeAr | 7.00 | 6.80 | 6.64 | 6.54 | 6.69 | 6.59 | 6.65 |
| NeAr | 6.94 | 6.79 | 6.69 | 6.63 | 6.71 | 6.66 | 6.65 |
| F$^-$He | 8.31 | 6.83 | 6.58 | 6.35 | 6.34 | 6.02 | 6.15 |
| F$^-$Ne | 8.81 | 6.28 | 6.34 | 6.19 | 6.09 | 5.90 | 5.94 |
| F$^-$Ar | 5.97 | 5.76 | 5.92 | 5.77 | 5.75 | 5.58 | 5.70 |
| Cl$^-$He | 8.58 | 7.88 | 7.82 | 7.43 | 7.65 | 7.16 | 7.51 |
| Cl$^-$Ne | 7.58 | 7.28 | 7.46 | 7.17 | 7.27 | 6.94 | 7.09 |
| Cl$^-$Ar | 7.00 | 6.89 | 7.05 | 6.87 | 6.89 | 6.73 | 6.96 |
| Li$^+$He | 3.76 | 3.69 | 3.60 | 3.59 | 3.63 | 3.60 | 3.58 |
| Li$^+$Ne | 4.02 | 3.95 | 3.90 | 3.88 | 3.90 | 3.88 | 3.86 |
| Li$^+$Ar | 4.58 | 4.53 | 4.51 | 4.49 | 4.52 | 4.38 | 4.47 |
| Na$^+$He | 4.56 | 4.49 | 4.37 | 4.40 | 4.46 | 4.49 | 4.39 |
| Na$^+$Ne | 4.80 | 4.74 | 4.65 | 4.68 | 4.69 | 4.73 | 4.66 |
| Na$^+$Ar | 5.36 | 5.30 | 5.19 | 5.28 | 5.24 | 5.32 | 5.25 |
| Li$^+$F$^-$ | 2.97 | 2.96 | 3.10 | 2.96 | 3.10 | 2.96 | 2.97 |
| Li$^+$Cl$^-$ | 3.87 | 3.85 | 3.95 | 3.83 | 3.88 | 3.87 | 3.84 |
| Na$^+$F$^-$ | 3.64 | 3.64 | 3.61 | 3.64 | 3.61 | 3.64 | 3.65 |
| Na$^+$Cl$^-$ | 4.50 | 4.48 | 4.40 | 4.47 | 4.44 | 4.51 | 4.48 |
| RMS | 0.84 | 0.22 | 0.18 | 0.09 | 0.08 | 0.11 | |
| RMS % | 13.8 | 3.8 | 2.9 | 1.4 | 1.6 | 1.7 | |

The abovementioned counterbalancing in the first and the second order interaction energies result in relatively small changes of the well depths for NeAr and Ar$_2$: as can be inferred from Table 2, D$_e$ changes by about ±2% for these two systems, whereas it systematically decreases by $5-10\%$ for the other rare gas dimers upon replacement of PBE0AC with ZMP. Not unsurprisingly, this is accompanied by a systematic lengthening of the equilibrium distance Re, which with ZMP increase by 0.02 bohr for NeAr and Ar$_2$ and by $0.04 - 0.07$ bohr for the other rare gas dimers.

Careful inspection of Table 2 reveals that the D$_e$ values obtained with ZMP-SAPT are closer to CCSD(T) than DFT-SAPT(PBE0AC) in three cases (He$_2$, HeNe, and HeAr) and farther away in two cases (Ne$_2$ and Ar$_2$), while for NeAr they yield similar agreement. The R$_e$ values of Table 3, on the other hand are improved upon replacing PBE0AC with ZMP in only two cases (He$_2$ and Ar$_2$) with respect to CCSD(T), and deteriorate in the remaining four dimers.

Table 4: % RMS errors of the interaction energies and minimum distances for the neutral dimers investigated. The reference methods for the first three rows was CCSD(T) at its respective basis set. All minima were determined with an Akima 1D Spline.

| Energies | PBE+$\delta(HF)$ | PBE0+$\delta(HF)$ | PBE0AC | PBE0AC+$\delta(HF)$ | ZMP | ZMP+$\delta(HF)$ | CCSD(T) |
|---|---|---|---|---|---|---|---|
| 5Z | 21.8 | 13.6 | 6.8 | 10.8 | 3.7 | 6.4 | Reference |
| 6Z | 18.4 | 13.4 | 7.3 | 12.2 | 3.9 | 7.4 | Reference |
| (5,6)Z | 20.7 | 12.7 | 7.7 | 12.7 | 3.8 | 7.6 | Reference |
| (5,6)Z | 21.0 | 13.1 | 7.9 | 12.9 | 4.0 | 7.8 | 1.9 |
| Distances | PBE+$\delta(HF)$ | PBE0+$\delta(HF)$ | PBE0AC | PBE0AC+$\delta(HF)$ | ZMP | ZMP+$\delta(HF)$ | CCSD(T) |
| 5Z | 6.4 | 3.4 | 0.65 | 1.19 | 0.72 | 0.83 | Reference |
| 6Z | 6.4 | 3.4 | 0.55 | 1.21 | 0.66 | 0.71 | Reference |
| (5,6)Z | 6.4 | 3.4 | 0.51 | 1.17 | 0.79 | 0.52 | Reference |
| (5,6)Z | 6.0 | 3.1 | 0.86 | 1.47 | 0.78 | 0.80 | 0.52 |

In Table 4, the basis set dependence of the RMS errors of the six rare gas dimers is investigated in more detail. We get a RMS error of 0.024 milliHartree (mH) for PBE0AC+$\delta(HF)$, 0.025 mH for ZMP+$\delta(HF)$, 0.012 mH for PBE0AC and 0.015 mH for ZMP when comparing to CCSD(T) at the aug-cc-pV(5,6)Z extrapolated basis set limit, confirming that the inclusion of $\delta(HF)$ is not favourable for these species. The comparison is valid, since CCSD(T) by itself deviates only by 0.003 mH RMS from the values from Tang-Toennies. The deviation of the energies from CCSD(T) does not change much with different basis set sizes, which is also reflected in the cbs limit. In Table 4, we show the % RMS errors since the interaction energies of the helium dimer and the argon dimer vary by more than an order of magnitude. The % RMS energy error of PBE0AC (7.9%) is about twice the RMS error of ZMP-SAPT (4.0%), which is in turn about twice the error of 1.9% for CCSD(T). ZMP+$\delta(HF)$ has an RMS error of 7.8%, respectively, followed by the standard PBE0AC+$\delta(HF)$, which has an error of 12.9%, i.e. more than three times that of ZMP without $\delta(HF)$. A similar picture emerges from Table 3 for the RMS errors of the minimum distances in comparison to those obtained from Coupled-Cluster: 0.073 Bohr (PBE0AC+$\delta(HF)$), 0.033 Bohr (ZMP+$\delta(HF)$), 0.032 Bohr (PBE0AC), and 0.050 Bohr (ZMP). For the distances, the deviation of CCSD(T) in comparison to the Tang-Toennies values is 0.031 Bohr, and the respective RMS errors for the SAPT values are 0.089 Bohr (PBE0AC+$\delta(HF)$), 0.048 Bohr (ZMP+$\delta(HF)$), 0.051 Bohr (PBE0AC), and 0.050 Bohr (ZMP). Since the minimum distances of the different dimers are not varying by a large amount, the RMS error and the % RMS error (Table 3) are much more closely correlated than in the case of interaction energies.

### D. Interaction energies for Dimers involving ions

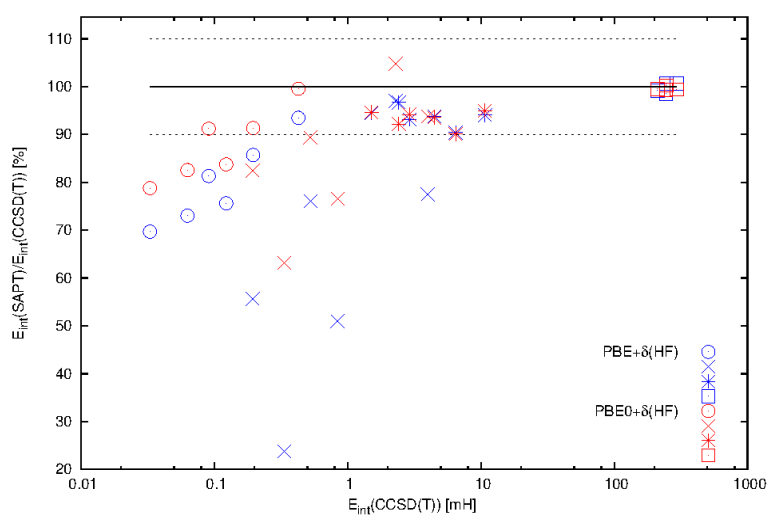

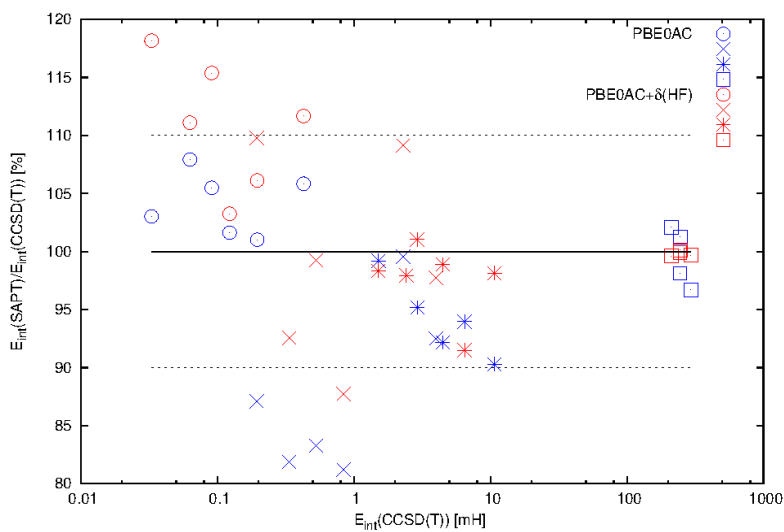

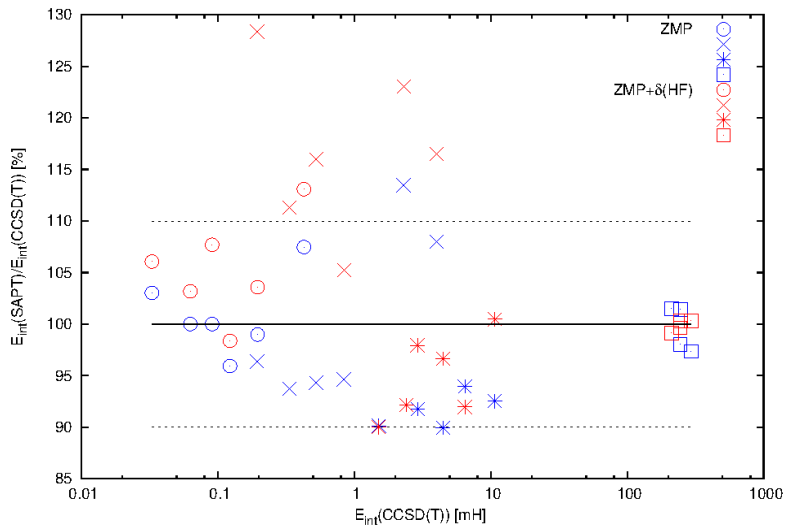

FIG. 5: Energy errors of DFT-SAPT including $\delta(HF)$ for the PBE and PBE0 xc potentials (a), of DFT-SAPT with and without $\delta(HF)$ for PBE0AC (b), and ZMP-SAPT with and without $\delta(HF)$ (c). The continuous line indicates the CCSD(T) benchmark, whereas the dotted lines are deviations of 10% from the reference. The neutral systems are indicated with circles, the anionic systems with an X, the cationic systems with a star, and the systems with two ions with a box. The only method which stays within these 10% limits is ZMP-SAPT without any correction, with the exception of Cl⁻Ar⁻.

Fig. 5 presents the ratios of DFT-SAPT well depths to CCSD(T) reference values in percentage form (based on the data of Table 2) for all considered dimeric species, i.e. also those containing ions.

The uppermost panel of Fig. 5 immediately makes clear that the hybrid xc potential PBE0 (red symbols) significantly improves upon the non-hybrid PBE (blue symbols). The latter in some cases dramatically underestimates the CCSD(T) well depth, even when the $\delta(HF)$ correction is taken into account which usually yields deeper wells (vide infra). Underestimation of the well depth by DFT-SAPT(PBE)+$\delta(HF)$ is particularly strong for F⁻He, where only 24% of the CCSD(T) well depth are reproduced, but is also very significant for the other anion-rare gas dimers. This is mainly due to the first-order exchange contribution: It is too repulsive with PBE (1.316 milliHartree (mH) for F⁻He at 6.2 bohr), since for local and gradient-corrected xc potentials electron densities are decaying too slowly. Incorporating 25% exact exchange in PBE0 yields a somewhat more rapidly decaying electron density and thus a lower $E_{exch}^{(1)}$ of 0.820 mH. Switching on the full asymptotic correction results in an even more rapidly decaying density with a correspondingly even less repulsive $E_{exch}^{(1)}$ of 0.616 mH for DFT-SAPT(PBE0AC) (for comparison, ZMP-SAPT yields 0.502 mH). For the cation-rare gas dimers, a similar effect on $E_{exch}^{(1)}$ can be observed, for example, for Li⁺He at 3.6 Bohr, $E_{exch}^{(1)}$ drops from 2.86 (PBE) over 2.53 (PBE0) to 2.11 mH (PBE0AC) (with ZMP 2.12 mH). This however, is partially compensated by the magnitude of the $E_{ind}^{(2)}$ term (-6.64, -6.18, -5.76, and -5.66 for PBE, PBE0, PBE0AC, and ZMP, respectively), so that for the cation-rare gas dimers the well depths hardy change upon replacing PBE with PBE0.

While seven systems with deviations of up to 40% and a general tendency to underestimate the well depths are found for PBE0, the maximum errors drop below 20% and only five systems lie systems outside the ± 10% error bars with PBE0AC. The RMS error is also reduced, i.e., from 0.62 to 0.30 mH, when applying the asymptotic correction, as is the percentage RMS error, from 13.0 to 7.7 % for all complexes (the latter is weighting the four ionic complexes less).

For the ionic complexes $Li^+F^-$, $Li^+Cl^-$, $Na^+F^-$, and $Na^+Cl^-$, it is important to use the estimate for the higher-order terms from $\delta(HF)$. These universially enlarge the interaction energy of the lithium-containing species and lower the interaction energy of the sodium containing species.

The middle panel of Fig. 5 demonstrates that inclusion of the $\delta(HF)$ correction in general leads to larger well depths, by up to 15%, with a few exceptions for cation-rare gas atom and cation-anion cases. Furthermore, when adding the $\delta(HF)$ correction to DFT-SAPT(PBE0AC) interaction energies (blue symbols) the well depths of the rare gas dimers in four out of six cases become overestimated by more than 10%, while addition of $\delta(HF)$ brings nearly all of the anion-rare gas dimers within the ± 10% error bars. The lowermost panel of Fig. 5 shows the corresponding comparison of pure ZMP-SAPT well depths (blue symbols) to their $\delta(HF)$-corrected counterparts (red symbols). Here we see that addition of the $\delta(HF)$ correction in general improves the well depths of cation-anion and cation-rare gas atom dimers, as it also was the case with the PBE0AC xc potential. However, as mentioned before, the $\delta(HF)$ correction is detrimental in case of the rare gas dimers. Though nearly all of their well depths stay within the ± 10% error bars, it significantly worsens the well depths of the anion-rare gas dimer systems, most of which show errors between 10 and 30% after addition of $\delta(HF)$. Without $\delta(HF)$, nearly all ZMP-SAPT well depths deviate by less than ± 10% from the CCSD(T) reference values.

When looking at ZMP-SAPT+$\delta(HF)$, we obtain somewhat worsened results than for PBE0AC. However, ZMP-SAPT shows the lowest % RMS error for all species when not including $\delta(HF)$, indicating that for very weakly bound systems, there is possibly some merit in neglecting this additional term. Thus, our best method for all 22 systems is not, as one may expect, PBE0AC+$\delta(HF)$ or ZMP+$\delta(HF)$, but rather ZMP without any additional high-order correction term.

For the minimum distances displayed in Table 3, the RMS errors are slightly smaller for ZMP when not including the additional $\delta(HF)$ extra term, and ZMP-SAPT performs slightly better than standard PBE0AC+$\delta(HF)$ compared to CCSD(T), at least when the standard RMS errors are concerned.

## IV. Conclusions

In summary, the results presented above show a promising strategy to improve DFT-SAPT which consists in replacing the exchange-correlation (xc) potential from a DFT model with a ZMP potential derived from *ab initio* electron densities, resulting in what conveniently can be named ZMP-SAPT. The first-order SAPT terms are directly affected by the replacement with the ZMP xc potential, which in the present study was generated from Brueckner Doubles calculations carried out with a triple-zeta quality basis set. Furthermore, the ALDA was employed to approximate the xc kernel required for the second-order SAPT contributions.

Please note that the ZMP xc potential for each monomer has to be generated only once to compute a complete ZMP-SAPT potential energy surface for interacting monomers (as long as the internal coordinates of the monomer do not change): this merely requires inexpensive translations/rotations of the grid points on which the xc potential is defined to the respective monomer positions or, preferentially, of its analytically fitted representation

Using up to augmented sextuple-zeta quality basis sets and complete basis set extrapolations, the well depths for the six rare gas dimers containing helium, neon, and argon atoms systematically improve with ZMP-SAPT over the best DFT-SAPT variant (which uses an asymptotically corrected PBE0 xc potential) in comparison to both, empirical Tang-Toennies and *ab initio* CCSD(T) potential energy curves. The overbinding tendency of DFT-SAPT(PBE0AC) for the rare gas dimers is practically eliminated with ZMP-SAPT, reducing the RMS percentage deviations from the CCSD(T) well depths from about 8% to 4%. This is mainly due to the first-order exchange-overlap contribution, which is systematically more repulsive with ZMP-SAPT. Adding the $\delta(HF)$ estimate of higher order SAPT contributions worsens the agreement of both, DFT-SAPT(PBE0AC) and ZMP-SAPT with the benchmark data for the considered rare gas dimers to about 13% and 8%.

In addition to these systems, which are weakly bound by dispersion interactions, we studied dimeric species containing anions and/or cations. For the strongly bound electrostatic interaction-dominated systems composed of lithium or sodium cations and fluorine or chlorine anions addition of the $\delta(HF)$ correction helps to further improve the well depths for both, DFT-SAPT(PBE0AC) and ZMP-SAPT. Here, however, even without the $\delta(HF)$ term both SAPT variants agree to better than 4% with CCSD(T). Both, DFT-SAPT(PBE0AC) and ZMP-SAPT with up to 10% deviate more strongly from CCSD(T) for the dimers composed of a cation and a rare gas atom, where the interactions are dominated by the induction contribution. Here, in most cases addition of $\delta(HF)$ somewhat improves the SAPT results with respect to the CCSD(T) well depths - hardly suprising in view of the fact that $\delta(HF)$ provides an estimate of higher-order induction and exchange-induction effects. The most demanding systems in our study are those composed of a rare gas atom and an anion, where electrostatic, induction, and dispersion interactions all are important for the overall binding. Four of the six considered dimers were significantly underbound (by more than 10%) with DFT-SAPT(PBE0AC) while with ZMP-SAPT only one dimer (Cl⁻Ar) deviates by more than 10% from CCSD(T), being overbound. Since the $\delta(HF)$ estimate increases the well depths for all anion-rare gas dimers, it seems to be beneficial for DFT-SAPT(PBE0AC) and detrimental for ZMP-SAPT.

Clearly, the ZMP-SAPT approach as used here could be improved in several respects. First of all, the computations of the electron densities required for construction of the ZMP xc potentials could be straightforwardly enhanced by using larger basis sets and even more accurate *ab initio* methods. Secondly, though the ALDA seems to work well for the xc kernel, a challenging goal would be to overcome it and to determine corresponding second-order SAPT contributions.[79] Thirdly, a better way to include higher-order SAPT contributions than estimating only part of them through the $\delta(HF)$ correction is needed. Third-order SAPT contributions have been derived and calculated for a few examples in the past,[80-83] and a Pauli-blockade approach which can be used to determine higher-order induction corrections on the DFT level recently has also been presented.[84] Furthermore, while the exchange-contributions of ZMP-SAPT are not potentially exact, as discussed for DFT-SAPT,[15] their quality in particular for short interatomic distances can be improved by avoiding the standard $S^2$-approximation for the second-order exchange terms.[85,86] Finally, it should not be forgotten that also the gold standard CCSD(T) is not free of approximations,[20,87] hampering the ultimate goal of reaching sub-wavenumber accuracy for weakly bound systems, as for example demonstrated for the case of the carbon monoxide dimer.[39]

The above number of approximations used in DFT-SAPT makes it difficult to improve DFT-SAPT systematically, something which is reminiscent of the problems associated with W2[88] and W3[89] theory. Here, only the computation of *all* missing contributions (HF basis set convergence, CCSD(T) correlation energy basis set convergence, post-CCSD(T) correlation, core-valence-correlation, diagonal Born-Oppenheimer correction, relativistic corrections) finally lead to some improvement resulting in the W4 method.[90] In the framework of single-determinant based SAPT approaches, the results presented above lead us to think that replacing a DFT model xc potential with an *ab* *initio* derived ZMP xc potential is one of the most important steps towards computation of spectroscopically accurate potential energy surfaces.